\documentclass[11pt,letterpaper]{article}
\usepackage[utf8]{inputenc}
\usepackage{amsmath}
\usepackage{amsfonts}
\usepackage{mathtools}
\usepackage{amssymb}
\usepackage{graphicx}
\graphicspath{{figures/}}
\usepackage{multicol}
\usepackage{authblk}
\usepackage[left=1.75cm,right=1.75cm,top=2cm,bottom=2cm]{geometry}
\usepackage{color}
\usepackage{placeins}

\author[1]{John W. Biddle\thanks{jbiddle@hcc-nd.edu}}
\affil[1]{Holy Cross College, Notre Dame, Indiana, USA}
\title{A method for calculating entropy production without knowing transition rates or dwell times}
\begin{document}
\maketitle
\begin{abstract}
The entropy production rate is central to the study of non-equilibrium systems.  This parameter is closely connected to violation of time-reversal symmetry, energy consumption, efficiency, and other properties of interest; in short, it quantifies how far a system is from thermodynamic equilibrium.  Standard formulas for the entropy production require knowledge of the system's underlying dynamics, but this knowledge may be hard to acquire in practice.  Here, I present a method for inferring the entropy production rate of a Markovian system from the \textit{sequence} of states that the system occupies on a single long trajectory or many shorter trajectories, and the total time of the trajectory or trajectories.  The method does not require knowledge of the dwell times in the various states, the times at which transitions occur, or of the transition rates characterizing the Markov process.
\end{abstract}

\section{Introduction}

Investigations into the physics of living organisms, in particular, are often concerned with the question of which processes are “thermal", i. e. taking place at thermodynamic equilibrium, and which ones are “active”, i. e., taking place away from thermodynamic equilibrium.  Investigators then, in turn, want to know just how ``active" the non-equilibrium processes are \cite{Li_Horowitz_2019,Fodor_2016, Gnessotto_Broedersz_2018}.  The non-equilibrium nature of a system or process can be thought of in a variety of ways, among them the violation of time-reversal symmetry, the consumption of energy, or the dissipation of heat.  Each of these can be quantified by means of the entropy production rate \cite{Maes_Netocny_2003,Otsubo_Krishnamurthy_2022}, which thus serves as the general parameter for departure from equilibrium \cite{Li_Horowitz_2019,Manikandan_Krishnamurthy_2021}.

If a system can be represented in terms of discrete, Markovian jumps between states, and all of the transition rates that characterize the dynamics of the system are known, calculating the entropy production rate for a non-equilibrium steady state (NESS) is straightforward \cite{Seifert_2012,Schnakenberg_1976}.  And improvements in experimental techniques over the past few decades have made single-molecule measurements possible and increasingly precise, greatly expanding the number of systems that can be treated in this way \cite{Ritort_2006,Miller_2018}. But for many systems of physical and biological interest, complete knowledge of the dynamics remains out of reach. In light of this, much research in the past several years has focused on methods for measuring or estimating various properties of non-equilibrium systems, in particular the entropy production rate, with incomplete information \cite{Seifert_2019,Fang_Wang_2019,Ertel_Seifert_2024}.  Several techniques have been proposed for Markov or semi-Markov processes in particular; Ref. \cite{Ertel_Seifert_2024} provides a good overview.

Methods that use coarse-grained representations of the system to deal with the reality of incomplete accessibility in the form of hidden or blurred states most typically rely on knowledge of certain transition rates or probability distributions \cite{Esposito_2012} or on information about dwell times in those states or collections of states that can be resolved \cite{Martinez_Parrondo_2019,Skinner_Dunkel_2021,Ertel_Seifert_2024}, or on the waiting-time distributions for particular transitions \cite{van_der_Meer_2022}.  Another family of methods relies on the observation of fluctuations over time of particular currents \cite{Li_Horowitz_2019,Horowitz_Gingrich_2020}. But it will not always be feasible to obtain dwell times, transition times, or the fluctuations of currents with the resolution needed to derive transition rates, find steady-state probabilities, or otherwise carry out the necessary calculations to find or estimate entropy production from these methods. Refs. \cite{Roldan_Parrondo_2010,Roldan_Parrondo_2012}, on the other hand take a different tack, similar to the method I present here, and use the relative probability of finding a sequence of states in forward and reverse order to detect departure from equilibrium and identify a lower bound on the entropy production. 

Building on recently published techniques for inferring the affinity of particular cycles in a Markov process \cite{Biddle_Gunawardena_2020,Pietzonka_Guioth_Jack_2021}, I present a method for calculating the entropy production rate of a Markovian system based on a single long trajectory, using only the sequence of states that the system visits, without regard to the time it spends in each state, and without any knowledge of either the transition rates or the steady-state probabilities that characterize the process.  This method converges to the exact entropy production in the long-time limit, and for certain system structures, it can also accommodate blurred or hidden states with no loss of accuracy.

\section{Definitions and Scope}

The results presented below apply to systems that can be modeled as stochastic processes occurring in continuous time with Markovian dynamics and a finite number of discrete microstates.  (Note that these need not be microstates in the original thermodynamic sense: see e. g. the Introduction to Ref. \cite{Esposito_2012}).  In general, these systems may be maintained away from thermodynamic equilibrium.  A tradition going back at least to Hill \cite{Hill_1966} and Schnackenberg \cite{Schnakenberg_1976} represents such systems with directed graphs.  This tradition is given a thorough exposition in Ref. \cite{Nam_2022}, and the notation in the present paper follows that of the ``linear framework" described therein.  The Markov process is represented by a graph $G$, of which the vertices $\{1,2,...,n\}$ represent the states of the system, the edges $i \to j$ represent possible transitions between states, and the edge labels $\ell(i \to j)$ represent the (Markovian) transition rates.  For present purposes the graph must be reversible, meaning if $i \to j$ exists, so does $j \to i$, corresponding to reversibility in the underlying dynamics.  The graph must also be strongly connected.  It will sometimes be convenient, when analyzing a directed graph $G$, to refer to the corresponding undirected graph, which will be denoted $\tilde{G}$.

A particular trajectory $X(t)$ of the underlying Markov process is specified by a sequence of states 
\begin{equation*}
\sigma_X = \{x_0, x_1, x_2, ... x_n\},
\end{equation*}
and a corresponding set of transition times,
\begin{equation*}
\tau_X = \{t_1, t_2, ... t_n\},
\end{equation*}
where $t_i$ is the time at which the system makes the transition into state $x_i$.

The transition rates/edge labels can be understood as a statement of conditional probability:
\begin{equation}
\ell (i\to j) = \lim_{\Delta t \to 0} \frac{\Pr(X(t+ \Delta t) = j | X(t) = i)}{\Delta t}.
\end{equation}
The edge labels of the graph can thus be used to construct a master equation for the evolution of microstate probabilities:
\begin{equation}\label{master}
\frac{dp_i}{dt}= \sum_{i \neq j} \big[ p_j \ell(j \to i) - p_i \ell (i \to j) \big],
\end{equation}
where $p_i(t)$ is the probability that the system is in state $i$ at time $t$.  The microstate probabilities can also be thought of as a stochastic column vector $\mathbf{p}(t)$ and Eq. \ref{master} as a linear differential equation
\begin{equation}
\frac{d \mathbf{p}}{dt} = \mathcal{L}(G) \cdot \mathbf{p}(t),
\end{equation}
where $\mathcal{L}(G)$ is the $n \times n$ Laplacian matrix of the graph $G$ \cite{Nam_2022}, commonly referred to as the ``rate matrix" or ``transition-rate matrix" of the system.  For a strongly connected graph whose transition rates are constant in time, $\mathbf{p}(t)$ approaches a unique steady state $\mathbf{p}^*$ \cite{Schnakenberg_1976, Nam_2022}.

\subsection{Illustrative example}

\begin{figure}[h!]
	\begin{center}
		\includegraphics[scale=0.6]{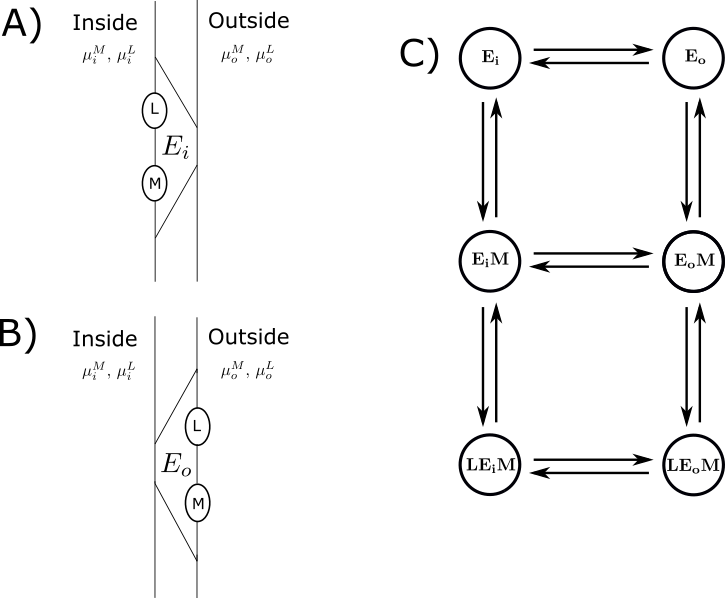}
	\end{center}
	\caption{A biological system used to illustrate the central result of the paper.  In this system, a protein complex E, which spans the cell membrane, facilitates active transport of a small molecule L against the chemical potential gradient. (A) is a schematic of the protein complex in conformation $E_i$, in which $E$ can only interact with small molecules $L$ and $M$ inside the cell membrane at chemical potentials $\mu_i^M$ and $\mu_i^L$. (B) Shows conformation $E_o$, in which the binding sites for $L$ and $M$ face the outside of the cell and binding of $L$ and $M$ take place at chemical potentials $\mu_o^M$ and $\mu_o^L$.  (C) shows a graphical representation of the possible states of $E$, also reflecting the limitation that $L$ can only bind if $M$ is also bound, and $M$ can only bind or unbind when $L$ is not bound.}
	\label{Fig-1-Example}
\end{figure}

This paper will often refer, as an example of a biological system that can be usefully treated as a Markov process unfolding on a graph, to a scenario treated in T. L. Hill's expository work, ``Free Energy Transduction and Biochemical Cycle Kinetics" (Ref. \cite{Hill_1989}).  In this scenario, a cell moves a small molecule $L$ from inside of the cell to outside of the cell, \textit{against} the chemical potential gradient.  It does this by coupling the transport to the movement of another small molecule, $M$, for which the chemical potential is greater inside of the cell than outside.  The coupling takes place by means of a large protein molecule $E$, which spans the cell membrane. $E$ has one binding site each for $L$ and $M$, and two conformations: in conformation $E_i$, the binding sites are accessible only to $L$ and $M$ molecules inside the cell; and in $E_o$ the binding sites are accessible only to $L$ and $M$ molecules outside the cell.  Conformational changes of $E$ are possible in any binding state, but $L$ is only capable of binding when $M$ is already bound, and $M$ is only capable of binding or unbinding when $L$ is not bound.  Fig. \ref{Fig-1-Example} shows the graph used to represent this system as a Markov process.

\section{Entropy Production in Markov Processes}

A system at thermodynamic equilibrium obeys the principle of detailed balance \cite{Ter_Haar_1954}: for Markov systems this means that the system has reached a steady state such that the net flux along each edge vanishes, meaning that each transition $i \to j$ is observed with the same frequency as the reverse transition $j \to i$ over any long trajectory. Here it is useful to define the net flux on an edge $i \leftrightarrow j$ at steady state as 
\begin{equation}
J_{ij} = p^*_i\ell(i \to j) - p^*_j\ell(j \to i) \label{fluxdef}
\end{equation}
If a Markov system is at thermodynamic equilibrium, it must be the case that 
\begin{equation}
J_{ij} = 0 \hspace{0.5 cm} \forall \hspace{0.2 cm} i\leftrightarrow j \in G, \label{db}
\end{equation}
that is, the net flux for all transitions must vanish at steady state \cite{Schnakenberg_1976}. 

Conversely, a NESS is a steady state for which not all net fluxes vanish.  A NESS is maintained by the coupling of the system to multiple heat baths, particle reservoirs, etc. at different temperatures, chemical potentials, etc. Whether or not a system can reach equilibrium, its Markovian dynamics is connected to the thermodynamics of the reservoirs through the principle of \textit{local} detailed balance, which obtains under conditions that are not very restrictive (e. g. the reservoirs must be good thermodynamic reservoirs) \cite{Seifert_2011,Bauer_Cornu_2015}.  Local detailed balance stipulates that (setting Boltzmann's constant to unity, as I will do throughout this paper),
\begin{equation}
\log \left[ \frac{\ell (i \to j)}{\ell (j \to i)}\right] = \Delta S^\mathrm{res} + \Delta S^\mathrm{sys}_{ij}, \label{ldb}
\end{equation}
with the right-hand side of Eq. \ref{ldb} representing the total change in the internal entropy of the system and the entropy of all reservoirs associated with a transition of the system from state $i$ to state $j$.  This yields one standard presentation \cite{Ertel_Seifert_2024} of the rate of entropy production $\dot{S}$ for a given NESS, summing the fluxes over all transitions, as follows:
\begin{equation}
\dot{S} = \sum_{i,j}p^*_i \ell(i \to j) \log \bigg(\frac{p^*_i \ell(i \to j)}{p^*_j \ell(j \to i)}\bigg). \label{entropy-transitions}
\end{equation}

Non-vanishing net fluxes at steady state are only possible in a graph that contains at least one cycle \cite{Hill_1966}.  Moreover, it follows from Eq. \ref{ldb} that for any cycle on the graph---that is, any sequence of transitions $\{i_1 \to i_2, i_2 \to i_3, \ldots, i_{m-1} \to i_m, i_m \to i_1\}$ that returns the system to its original state---we can identify the thermodynamic affinity of the cycle,
\begin{equation}
\tilde{A}(C) = \log \left[ \frac{\ell (i_1 \to i_2) \times \ell (i_2 \to i_3) \times \cdots \times \ell(i_m \to i_1) }{\ell (i_1 \to i_m) \times \ell (i_m \to i_{m-1}) \times \cdots \times \ell(i_2 \to i_1)}\right], \label{affinity}
\end{equation}
with the total entropy produced in the reservoirs each time the cycle is traversed.  We define $C^r$, which we will also refer to as the reverse cycle of $C$, as a cycle consisting of the same states, but occurring in the opposite order. (Note that this means that $C^r$ also has the same initial state as $C$.)

The association of Eq. \ref{affinity} with entropy has led to several expressions for entropy production that are equivalent to Eq. \ref{entropy-transitions}, and are based on cycles rather than individual transitions.  Of particular interest here is Schnackenberg's method, which uses only a subset of the cycles on the graph, called a fundamental set of cycles, and counts the transitions on representative edges rather than completions of cycles.  The central result of this paper builds on Schnakenberg's method \cite{Schnakenberg_1976}, but Schnakenberg's formula required the affinities to be calculated from transition rates, and therefore required that the transition rates themselves somehow be known; the method presented here does not require such knowledge.  Schnakenberg's method is summarized in the next section, and the central result of this paper in the following section.  Other examples of expressions for entropy production based on the properties of cycles include the work of Hill and Simmons \cite{Hill_Simmons_1976} and of Jiang \textit{et al.} \cite{JiangDa-Quan2004Mton}, both of which introduced expressions for entropy production.  The expressions developed in the latter can also be used to infer entropy production without knowing transition times, but they lack the flexibility of a method based on Schnakenberg's decomposition into fundamental cycles.  

\subsection{Schnakenberg's entropy production}

\begin{figure}[h!]
	\begin{center}
		\includegraphics[scale=0.6]{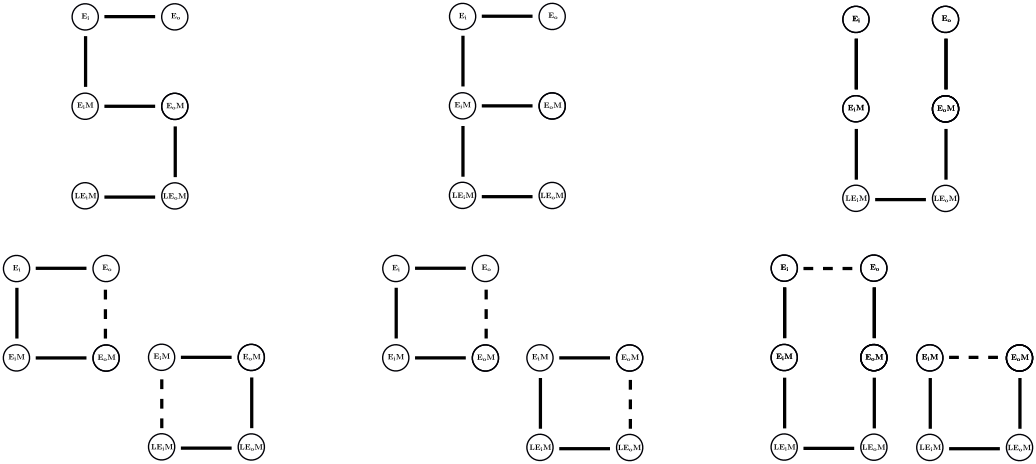}
	\end{center}
	\caption{Top row: spanning trees of the graph in Fig. \ref{Fig-1-Example}  Bottom row: the fundamental set of cycles associated with each spanning tree; the chords associated with each cycle are shown as dashed rather than solid lines.}
	\label{Fig-2-Trees}
\end{figure}

The first step of Schnakenberg's method is to decompose a graph into a set of fundamental cycles.  Such a set can be found by starting with a ``spanning tree" of the undirected graph $\tilde{G}$.  A spanning tree $\tilde{T}(\tilde{G})$ is a subgraph of $\tilde{G}$ that contains all of the vertices of $\tilde{G}$, but only a subset of the edges, so that $\tilde{T}$ remains strongly connected but contains no cycles.  The edges of $\tilde{G}$ that were removed to create $\tilde{T}$ are called ``chords".  A set of fundamental cycles is then constructed as follows: the chords of $\tilde{T}(\tilde{G})$ are restored, one by one, to $\tilde{T}$.  The first chord to be restored creates one cycle. (In translating this to a cycle on the graph $G$, a direction must be chosen for the cycle, but the choice is arbitrary and does not affect any of our results). The addition of each subsequent chord will create exactly one cycle that does not contain any other chords, and this cycle is included in the fundamental set of cycles; the addition of a chord may also create additional cycles that contain previously added chords as well, but these are excluded from the fundamental set of cycles. We define the net flux on the chord $I^\alpha$ associated with  cycle $C_\alpha$ as $J^\alpha = p^*_i\ell (i^\alpha  \to j^\alpha) - p^*_j \ell (j^\alpha \to i^\alpha)$, where $i^\alpha$ and $j^\alpha$ are the vertices connected by $I^\alpha$, with $i^\alpha$ being the one that occurs first in the direction specified for the cycle.  Fig. \ref{Fig-2-Trees} contains some examples of spanning trees and corresponding sets of fundamental cycles for the graph introduced in Fig. \ref{Fig-1-Example}.

Schnakenberg proved that the entropy production rate can then be represented in terms of the affinity of each of the fundamental cycles and the net flux at steady state on the chord associated with that cycle:
\begin{equation}
\dot{S} = \sum_\alpha \tilde{A}(C_\alpha) J^\alpha, \label{schnakenberg}
\end{equation}
and Eq. \ref{schnakenberg} is valid for any choice of fundamental cycles \cite{Schnakenberg_1976}.

In Schnakenberg's formulation, the affinities of the cycles can be calculated directly from Eq. \ref{affinity} if the transition rates are known.  They can also be deduced from knowledge of the thermodynamic forces on the system being modeled; in the example system provided, they could be deduced if one knew the chemical potentials of $L$ and $M$ both inside and outside of the cell.  A recent result provides a method of inferring information about the affinities of cycles from waiting-time distributions of transitions on particular edges \cite{van_der_Meer_2022} . And because spanning trees also provide a means of calculating the steady-state probabilities of each state in any strongly connected graph from the transition rates of that graph, the net fluxes can also be calculated from the transition rates if they are all known.

However, as discussed in the Introduction, transition rates may not be known, the information on dwell times may not be sufficient to resolve them, and the thermodynamic forces on the system may not be independently known.

\section{Central Result}
A cycle $C_\alpha$ specifies a sequence of states that the Markov process occupies.  In a trajectory $X(t)$, we can let $n[C_\alpha, X(t')]$ represent the number of times that the sequence of states specified by $C_\alpha$ occurs (exactly, in order) in the sequence $\sigma_X$ up to time $t'$, and let $n[C^r_\alpha, X(t')]$ represent the number of times that cycle $C_\alpha$ occurs in reverse order in the same sequence.  The essential contribution of Ref. \cite{Biddle_Gunawardena_2020} is that for simple cycles, \textit{i. e.}, cycles that do not repeat a transition,
\begin{equation}
\lim_{t \to \infty} \frac{n[C_\alpha,X(t)]}{n[C^r_\alpha,X(t)]} = e^{\tilde{A}(C_\alpha)}. \label{BG-Result}
\end{equation}
Ref. \cite{Pietzonka_Guioth_Jack_2021} extended this result in two significant ways (among others).  First, it showed that a long trajectory is not necessary: if one averages over many trajectories of finite length $\tau$, one finds
\begin{equation}
 \frac{\left\langle n[C_\alpha,X(\tau)]\right\rangle}{\left\langle n[C^r_\alpha,X(\tau)]\right\rangle } = e^{\tilde{A}(C_\alpha)}, \label{PGJ-Generalization}
\end{equation}
where the angle brackets represent an average over trajectories.  Second, it showed that Eqs. \ref{BG-Result} and \ref{PGJ-Generalization} need not be limited to simple cycles.  Indeed, they also hold more generally for ``families" of cycles, where a family of cycles is any set of cycles sharing the same affinity.  (The simplest example is that cycles sharing the same vertices in the same order, but with a different initial vertex, form a family of cycles).  Let $\mathcal{F}$ represent a family of cycles under this definition, and let $n[\mathcal{F},X(t')]$ represent the total number of occurrences of all cycles in a family $\mathcal{F}$ up to time $t'$.  Then let $\mathcal{F}^r$ refer to a set consisting of the reverse cycle $C^r$ of each cycle $C\in\mathcal{F}$.  Then it is also the case that,
\begin{equation}
 \frac{\left\langle n[\mathcal{F},X(\tau)]\right\rangle}{\left\langle n[\mathcal{F}^r,X(\tau)]\right\rangle } = e^{\tilde{A}(C_\alpha)}, \label{PGJ-GeneralizationII}
\end{equation}
and 
\begin{equation}
\lim_{t \to \infty} \frac{n[\mathcal{F},X(t)]}{n[\mathcal{F}^r,X(t)]} = e^{\tilde{A}(C_\alpha)}. \label{PGJ-GeneralizationIII}
\end{equation}

Note that these families need not be limited to simple cycles: vertices and transitions can be repeated, as long as the initial vertex is not repeated more than once.  \textit{A fortiori}, Eq. \ref{BG-Result} need not be limited to simple cycles.

Let  $n[i \to j, X(t')]$ represent the number of times that the sequence of states $i,j$ occurs in $\sigma_X$ up to time $t'$.  The ergodic property of strongly-connected Markov systems implies a relationship between $n[i \to j, X(t')]$ and the steady-state flux that has been useful to in thermodynamic inference \cite{Skinner_Dunkel_2021b,Nitzan_2023}:
\begin{equation}
\lim_{t \to \infty} \frac{n[i \to j, X(t)]-n[j \to i, X(t)]}{t} = p^*_i\ell(i \to j) - p^*_j\ell(j \to i) \label{Ergodic-Flux}.
\end{equation}

Eqs. \ref{BG-Result} and Eq. \ref{Ergodic-Flux} allow Schnakenberg's entropy production to be expressed entirely in terms that depend only on the \textit{sequence}  $\sigma_X$ and the \textit{total time} of a single long trajectory, as
\begin{equation}
\dot{S} =\lim_{t \to \infty} \sum_\alpha \log \Bigg( \frac{n[C_\alpha,X(t)]}{n[C^r_\alpha,X(t)]} \Bigg) \times  \frac{n[i^\alpha \to j^\alpha, X(t)]-n[j^\alpha \to i^\alpha, X(t)]}{t} \label{Result}
\end{equation}

Nothing in this expression requires any information about the times at which transitions occurred, nor is it necessary to infer anything quantitative about the underlying transition rates.  It is thus possible to infer the entropy production rate of the system without any information about transition rates, transition times, or dwell times.  All that is required is the sequence of states $\sigma_X$ and the total time of the trajectory.

\subsection{Considerations of hidden states and coarse-graining}

In addition to providing an expression for the entropy production rate that does not depend on transition rates or dwell times, Eq. \ref{Result} can, in some cases, be used in conjunction with Eqs. \ref{PGJ-GeneralizationII} and \ref{PGJ-GeneralizationIII} to obtain the entropy production without any loss of accuracy for partially accessible Markov states with blurred transitions.  

Recall that in order to compute the entropy production, one needs only to (1) ascertain the affinity of one set of fundamental cycles, and (2) calculate the net flux at steady state on the chords of those cycles.  In order to ascertain the affinity of a fundamental, simple cycle $C^\alpha$, one need not be able to resolve particular instances of that $C^\alpha$. It is sufficient to be able to resolve instances of some family $\mathcal{F}\ni C^\alpha$ and of $\mathcal{F}^r$.

\begin{figure}[h!]
	\begin{center}
		\includegraphics[scale=0.4]{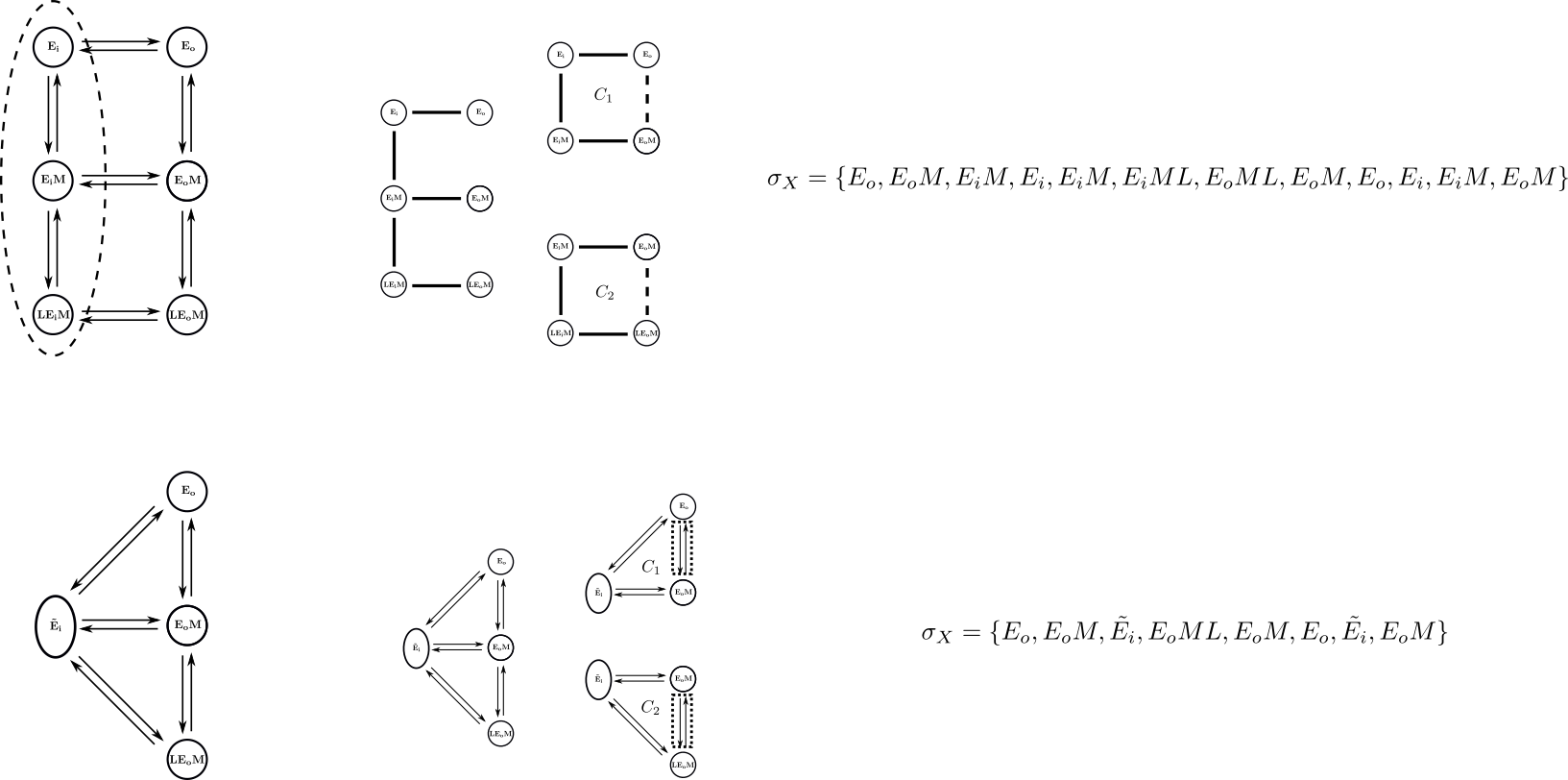}
	\end{center}
	\caption{Top: The system described in Fig. \ref{Fig-1-Example}, together with a decomposition into fundamental cycles and a hypothetical sequence $\sigma_X$ associated with a possible trajectory $X$.  Bottom: a coarse graining of the same system corresponding to the binding state being unobservable if the protein complex is in conformation $E_i$.  A decomposition of this coarse-grained system into fundamental cycles is also shown, along with the blurred sequence that would be recorded by an observer subject to this limitation.}
	\label{Fig-Coarse-Graining}
\end{figure}

Consider a system like that described in the Illustrative Example, but in which the binding state cannot be known, nor can binding and unbinding transitions be resolved, if the complex is in conformation $E_i$.  An observer would not be able to record the sequence $\sigma_X$ for an observed trajectory; the observer would only have access to a coarse-grained sequence, which we will call $\sigma'_X$, playing out on a coarse-grained version of the graph $G$, which we will call $G'$.  This is illustrated in Fig. \ref{Fig-Coarse-Graining}.  One would then merely have to choose a set of fundamental cycles such that these binding and unbinding transitions are not chords, and all of the necessary information will still be available. 

\subsection{Convergence and Shorter Trajectories}
Eq. \ref{Result} suggests an empirical estimator for the entropy production rate that converges in the long-time limit:
\begin{equation}
\hat{\dot{S}}(t) = \sum_\alpha \log \Bigg( \frac{n[C_\alpha,X(t)]}{n[C^r_\alpha,X(t)]} \Bigg) \times  \hat{J}^\alpha (t), \label{Estimator}
\end{equation}
where
\begin{equation}
\hat{J}^\alpha (t) = \frac{n[i^\alpha \to j^\alpha, X(t)]-n[j^\alpha \to i^\alpha, X(t)]}{t}.
\end{equation}
It is difficult, however, to make general statements about the time, or the number of transitions, needed for Eq. \ref{Estimator} to converge. In fact, the various cycle affinities and fluxes in a given system may differ by an order of magnitude or more in how long it takes them to converge satisfactorily, and so the time over which Eq. \ref{Estimator} converges to the true steady-state value of the entropy production rate may depend a great deal on the choice of fundamental cycles and representative edges.  This is one of the strengths of the result presented here: since one has the freedom to choose any set of fundamental cycles, one can choose to use the set of fundamental cycles that yields the fastest convergence.  This is elucidated below in the section ``Illustration."

Eq. \ref{PGJ-GeneralizationII} (first presented in Ref. \cite{Pietzonka_Guioth_Jack_2021}) can also be leveraged to increase the utility of the central result and provides a path for inferring entropy production from many shorter trajectories rather than one long one.  Eq. \ref{PGJ-GeneralizationII} is specific to cycles and families of cycles, and does not apply to the net flux on an edge, so an average of Eq. \ref{Estimator} evaluated over several short trajectories will not necessarily converge to the correct value of the entropy production.  Still, the trajectories need not be long enough for Eq. \ref{Estimator} to converge to its long-time limit: only the fluxes on the representative edges for the chosen set of fundamental cycles need to converge. The corresponding cycle affinities can be calculated by averaging over many such trajectories.

\section{Illustration}
\begin{figure}[!h]
\begin{center}
		\includegraphics[scale=0.8]{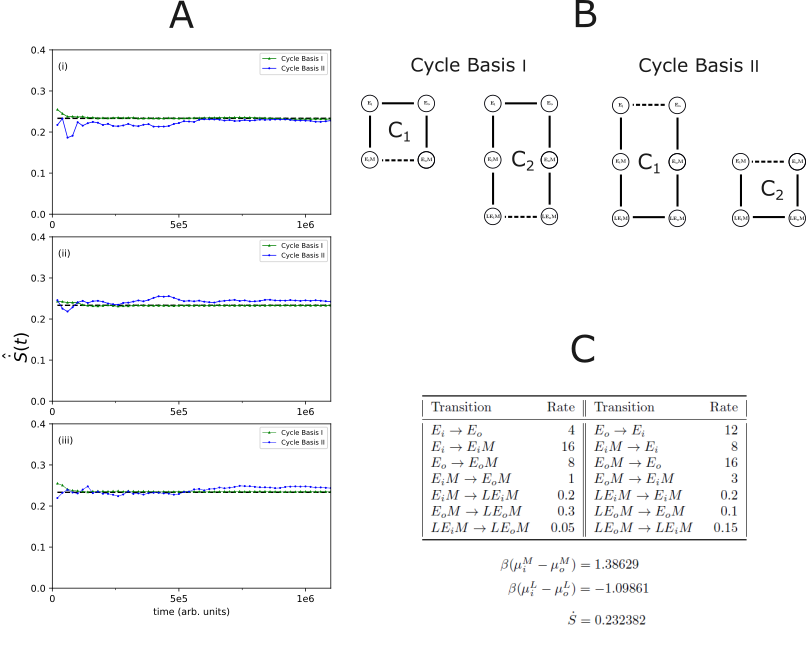}
	\end{center}
	\caption{Results of Gillespie simulations modeling the system described in the Illustrative Example.  (A) The finite-time estimator for the entropy production rate, $\hat{\dot{S}}(t)$ as defined in Eq. \ref{Estimator}, evaluated at regular time intervals using two different cycle bases and compared to the entropy production rate calculated using Eq. \ref{schnakenberg} (shown as a black dashed line). For these calculations,  $k_B$ is set to 1 and the units of time are arbitrary, normalized so that $\ell(E_iM \to E_oM) = 1$. $\hat{\dot{S}}(t)$ is calculated from the elapsed time and the sequence $\sigma_X$, not from the parameters of the model itself.  (B) The cycle bases used to calculate $\hat{S}(t)$. (C) The parameters of the model and the entropy production rate calculated from those parameters.}
	\label{Fig-Illustration}
\end{figure}

To illustrate this result, I have carried out Gillespie simulations on the system described in the Illustrative Example.  The system is simulated with transition rates chosen to be consistent with $\mu^M_i - \mu^M_O > \mu^L_O - \mu^L_i$ and normalized so that $E_i M \to E_O M = 1$.  The (normalized) entropy production rate is calculated explicitly, as a comparison.  Then, for three trajectories of the system, an analysis is presented based only on the sequence of states against a ``running clock": the total time of the trajectory is included in the analysis, but transition times are not.  The entropy production rate is evaluated at successive times (always with $\Delta t \gg 1$) using Eq. \ref{Estimator}.  The entropy production rate is calculated for each trajectory with two different choices of fundamental-cycle decomposition to highlight the benefit of an advantageous choice of fundamental cycles and representative edges enabled by the central result of this paper.  While Cycle Basis II (as defined in Fig. \ref{Fig-Illustration}) does not converge within the time of the simulation study, Cycle Basis I converges quickly and satisfactorily for all three simulated trajectories.

\section{Discussion}

The entropy production rate is the most concise and informative parameter pertaining to a system’s departure from equilibrium.  As such, it continues to be a topic of active interest to researchers who study non-equilibrium systems, biological and otherwise.  The work in this paper draws on foundational work in stochastic thermodynamics \cite{Schnakenberg_1976} with much more recent results pertaining to thermodynamic inference \cite{Biddle_Gunawardena_2020,Pietzonka_Guioth_Jack_2021} to arrive at a way of obtaining the entropy production rate of an observed system with Markovian dynamics--a way that does not depend on knowing or even inferring transition times, dwell times, or transition rates.  In particular, the sequence of states that a system visits is sufficient to construct an estimator that converges, in the long-time limit, to the entropy production rate.

Recent experimental developments, such as single-particle tracking experiments, have made it possible to observe biological systems in particular in much greater detail; yet there is much that they still cannot observe or can only observe partially or with a great deal of uncertainty.  Consequently, much recent work in statistical physics and the physics of living systems has focused on inferring such properties as the entropy production rate from incomplete information.

One obstacle to inferring the entropy production rate of a particular system is the difficulty of inferring the underlying transition rates that characterize it.  The results presented in this paper obviate this concern by providing a way to calculate the entropy production rate only from the sequence of states that the system occupies, without needing any information about dwell times or transition rates.  These expressions also permit the inference of the entropy production in a way that is unaffected by the blurring of certain transitions and the inability to distinguish certain states, though the details of this aspect of the result depend on the structural details of the graph that can be used to represent the system.  This work derives additional flexibility and utility from the fact that it applies to any fundamental set of cycles. It can therefore be implemented on the basis of data pertaining only to a subset of cycles.  The convergence rate of the estimator presented here can vary greatly depending on the details of the system in question, and additional work on the question of convergence, applied to particular systems of interest, would be valuable.  Moreover, since the convergence times of different cycles and fluxes even within a particular system can be dramatically different, the ability to choose a cycle basis to one's advantage can be very useful.  

The results presented here contribute to a growing body of work to help researchers quantify departure from equilibrium despite limitations on their ability to observe the system.

\FloatBarrier
\section*{Acknowledgements}
The author thanks William Pi\~{n}eros for a seminal suggestion that led to this investigation, and he also thanks Jeremy A. Owen and Alex Albaugh for useful discussions and feedback on an early draft of the manuscript.

\section*{Declaration of generative AI and AI-assisted technologies in the manuscript preparation process}
The author used ChatGPT to de-bug and improve the python code for the Gillespie simulations reported in the manuscript; and to proof-read the manuscript for spelling, grammar, syntax, and style and to evaluate its overall coherence and suitability.  The author himself wrote and edited the text of the manuscript and is fully responsible for its contents.


\end{document}